# Robust substrate anchorages of silk lines with extensible nano-fibres

Jonas O. Wolff[1*,a], Daniele Liprandi[2,a], Federico Bosia[3], Anna-Christin Joel[4], and Nicola M. Pugno[2,5,**]

[1] Department of Biological Sciences, Macquarie University, Sydney, NSW 2109, Australia

[2] Laboratory of Bio-inspired, Bionic, Nano, Meta Materials & Mechanics, Department of Civil, Environmental and Mechanical Engineering, University of Trento, Via Mesiano 77, I-38123 Trento, Italy

[3] Department of Applied Science and Technology, Politecnico di Torino, Corso Duca degli Abruzzi 24, 10129, Torino, Italy

[4] Institute of Biology II, RWTH Aachen University, Worringerweg 3, 52074 Aachen

[5] School of Engineering and Materials Science, Queen Mary University, Mile End Rd, London E1 4NS, UK

[*] corresponding author: jonas.wolff@mq.edu.au
[**] corresponding author: nicola.pugno@unitn.it
[a] These authors contributed equally to this work.

**Abstract**

Living systems are built of multiscale-composites: materials formed of components with different properties that are assembled in complex micro- and nano-structures. Such biological multiscale-composites often show outstanding physical properties that are unachieved by artificial materials. A major scientific goal is thus to understand the assembly processes and the relationship between structure and function in order to reproduce them in a new generation of biomimetic high-performance materials. Here, we tested how the assembly of spider silk nano-fibres (i.e. glue coated 0.5μm thick fibres produced by so-called piriform glands) into different micro-structures correlates with mechanical performance by empirically and numerically exploring the mechanical behaviour of line anchors in an orb weaver, a hunting spider and two ancient web builders. We demonstrate that the anchors of orb weavers exhibit outstanding mechanical robustness with minimal material use by the indirect attachment of the silk line to the substrate through a soft domain ('bridge'). This principle can be used to design new artificial high-performance attachment systems.

## 1. Introduction

The materials that living systems are built of and produce often display mechanical properties that outperform most man-made materials. This is largely attributed to the specific nano- and micro-composite structure of such materials: by the combination of components with distinct properties and their specific spatial arrangement, the resulting material exhibits a mix of characteristics that is usually not possible to achieve with independent phases or homogeneous (and isotropic) bulk materials (1). Biological composites are also common among adhesive secretions and organs used by organisms to achieve substrate attachment, such as in spider silk anchors (2), mussel byssus threads (3) and gecko feet (4). This is in contrast to artificial adhesive joints that are usually based on homogeneous bulk materials. A better understanding of the structure-function relationships in biological structured adhesives could inspire a new generation of adhesive systems and fasteners with previously unachieved properties (5, 6).

Spider web anchors are micro-structures that serve as attachments for silk lines. Unlike most other biological composite structures and materials, such as bone, nacre or wood, these micro-structures are rapidly formed by the instant spinning of fibres from a secretion stock and their behavioural assembly into distinct architectures. As the fibre arrangement affects the mechanical properties of the structure, a spider can tailor the end-product for different needs using the same base material (7). These features make spider silk anchors very interesting for the bio-inspiration to design instant attachments with tailored properties. However, a successful implementation into a bio-inspired design is still hindered by a limited understanding of the functional role of the material mix and spatial distribution of mechanical properties in silk anchorages.

Silk anchors attach a structural thread, the dragline, to the substrate by a plaque (membrane) that is composed of numerous adhesive nano-fibres (diameter ~0.5 μm). The base material of the plaque is the so-called *piriform silk*. Tensile stress tests of this special type of silk have revealed that it is three times more extensible than the common dragline silk (*major ampullate silk*) (8, 9).

Previous research has shown that the morphology of spider web anchors can be approximated as a tape-like membrane with the dragline fused along the central axis (10). The location where the dragline exits the membrane ('loading point') thereby affects the maximal force the anchor can sustain under tensile load (10). If the dragline is stressed, the membrane is locally deformed and detaches from the substrate surface, along what is referred to as "peeling line". Membrane delamination thus leads to a changing -often growing- peeling line. After

Kendall's peeling theory (11) the peeling force of elastic films emerges as proportional to the length of the peeling line. It can be empirically observed, that if the anchor loading point is close to the membrane edge, the peeling line is roughly V-shaped, whereas it is concentric when the loading point is in a central position developing a larger length before meeting the membrane edge (10). Previous comparative measurements and numerical simulations uncovered that dragline placement modulates the anchor's strength, with the 'centrality' of the dragline joint $c_d$ (i.e. the distance between the loading point and the front edge of the anchor membrane divided by the membrane length) being the main determinant of anchor performance within species (10, 12).

It remains uncertain how the differences in mechanical properties of the attachment silk and the dragline affect anchor strength, and if there is an optimal ratio of extensible and stiff materials combined in the anchorage. Also, explorative studies have shown that the structure of the dragline joint differs between species, with the dragline being either directly embedded in the piriform silk film in some or suspended in a flexible network of piriform fibres (called 'bridge') in others (13) (Fig. 1). The bridge is built of the same material as the silk film applied to the substrate (the plaque). However, while in the plaque the piriform fibres are arranged in a lattice-like structure, they form a radial network of branching bundles in the bridge (Figs. 1h, p). This may reduce stress concentrations in the loading point and enhance anchor resistance at steep pulling angles. A previous in-depth study on the anchorages of the golden orb weaver *Trichonephila plumipes* revealed that the bridge was asymmetric with a stronger expression in the upstream (i.e. the leaving) dragline insertion (10). When pulling on the upstream dragline, a higher force was required to detach the anchor from a polymer surface than when pulling on the downstream dragline, were the bridge was much smaller (10). Therefore, we hypothesized that the bridge enhances the overall robustness of the anchorage. To test this conjecture, we combined an empirical and a numerical study of anchor detachment mechanics in four species that differ in the structure of the dragline joint.

## 2. Results

*2.1. Effects of structure on silk anchor mechanics*

For the comparative analysis of anchor mechanics, four species were chosen that differed in their silk ecology and silk anchor structure. The Southern house spider *Kukulcania hibernalis* (Fig. 1a) and the Tasmanian cave spider *Hickmania troglodytes* (Fig. 1b) are ancient sheet web builders that barely changed over a long evolutionary timescale. Both species build sheet webs, but while in *K. hibernalis* the sheet is placed directly on the substrate surface, the sheet of *H. troglodytes* is horizontally suspended between rocks which may mechanically be more demanding. The huntsman *Isopeda villosa* (Fig. 1c) is an arboreal hunting spider that uses dragline to secure itself but does not build webs for prey capture. The golden orb weaver *Trichonephila plumipes* (Fig. 1d) is a large aerial web builder and represents the most advanced silk use.

We found that in *K. hibernalis* the dragline was directly incorporated into the membrane, with single strands separately following the looping patterns of the piriform silk fibres (Figs. 1e, j, r). In the three other species, the dragline fibres were bundled and did not follow the looping patterns of piriform silk fibres in the membrane but were oriented along the middle axis of the more or less axial-symmetrical membrane. While the dragline bundle was attached on top of the membrane in *H. troglodytes* (Figs. 1f, l, s), it was suspended in a radial network of piriform fibre bundles (called piriform silk 'bridge') in *T. plumipes* (Figs. 1h, p, u; see also 3D-reconstruction and sections in Wolff & Herberstein (10)). A piriform silk bridge was also present in *I. villosa*, but to a lesser extend (Figs. 1g, n, t).

We compared the nominal anchor strength $F_{max}/A$ of the anchors of the four species as the maximal pull off force divided by the plaque area, by pulling off the samples perpendicularly from a (here polypropylene) surface. Nominal strength differed between all species (*K. hibernalis* – *H. troglodytes*: $p=0.037$; *K. hibernalis* – *I. villosa*: $p<0.001$; *K. hibernalis* – *T. plumipes*: $p<0.001$; *H. troglodytes* – *I. villosa*: $p=0.030$; *H. troglodytes* – *T. plumipes*: $p<0.001$; *I. villosa* – *T. plumipes*: $p=0.015$). It was higher in anchors with a bridge than in those without a bridge. Anchors of *T. plumipes* yielded a ten times higher strength than the anchors of *K. hibernalis*, on average (Fig. 2a). We further analysed the relationship between $c_d$ and $F_{max}$. Due to the intraspecific variation of silk anchors, we could obtain and measure samples covering a range of $c_d$ values. In all species, $F_{max}$ was a clear function of $c_d$ (*K. hibernalis*: $R^2 = 0.554$, $p = 0.006$; *H. troglodytes*: $R^2 = 0.550$, $p = 0.001$; *I. villosa*: $R^2 = 0.576$, $p < 0.001$; *T. plumipes*: $R^2 = 0.624$, $p < 0.001$). However, assuming a linear relation $F_{max}/A = a \cdot c_d$, slopes varied between species from low in *K. hibernalis* ($a = 6.97$ mN/mm$^2$) to steep in *T. plumipes* ($a = 34.28$ mN/mm$^2$) (Fig. 2b).

In addition to perpendicular ($\theta = 90°$) pull-off tests, we performed similar tests at $\theta = 0°$ (i.e. along the dragline, parallel to the substrate) and $\theta = 180°$ (dragline flipped over) in all species except *K. hibernalis*. We found that $F_{max}$ was highest at $\theta = 0°$ in all species. However, the reduction of anchor resistance at $\theta = 90°$ and $\theta = 180°$ differed between species (Figs. 3b, d, f). Notably, $F_{max}$ was negatively proportional to the pull-off angle in *H. troglodytes* (0° vs. 90°: $p<0.001$; 0° vs. 180°: $p<0.001$; 90° vs. 180°: $p=0.909$) (Figs. 3a, b), while force drops were highest at $\theta = 90°$ in *I. villosa* (0° vs. 90°: $p<0.001$; 0° vs. 180°: $p=0.001$; 90° vs. 180°: $p=0.001$) (Figs. 3 c, d) and *T. plumipes* (0° vs. 90°: $p<0.001$; 0° vs. 180°: $p<0.900$; 90° vs. 180°: $p<0.001$) (Figs. 3 e, f). Force-distance plots indicated that initial stiffness values of anchor joints were similar in all three loading situations for *H. troglodytes* (Fig. 3a) and *I. villosa* (Fig. 3c). However, in *T. plumipes* increasing pull-off angles led to decreased anchor stiffness and higher deformation of the silk joint (Fig. 3e). Furthermore, the proportion of silk anchors that failed by internal fracture before substrate detachment were lowest in *T. plumipes* (pie charts in Fig. 3f), indicating that here interfacial stress concentrations were prevented by structural means and in general that anchors work roughly around the uniform strength for peeling and fracture (optimal) condition. A closer inspection of the force-distance curved revealed that the initial slopes (indicating the structural stiffness) were more or less similar in all loading situations in both *H. troglodytes* and *I. villosa*, but decreased with an increasing loading angle in *T. plumipes*. This indicates that stress level is mitigated by structures with different stiffness in the latter species, with a greater concentration on a stiffer structure at low angles and a greater concentration on a softer structure at steep angles. To rationalize these complex observations, we performed ad hoc *in silico* experiments.

*2.2. Model properties*

Numerical simulations can help to clarify the mechanisms of anchor detachment and investigate the influence of the mechanical and geometrical parameters, as the possibility of experimental manipulation is limited. We developed a numerical approach for the three-dimensional deformation and detachment of an elastic membrane (approximating the silk anchorages) of variable geometry and elastic properties, modelled as a spring lattice adhering to a rigid substrate through a cohesive interface (14). Spider-silk is a non-linear elastic material characterized by a yielding-hardening stress-strain law (8). However, as documented in the Supplementary Material, by implementing a cubic law stress-strain law, there are limited variations of the maximal pull-off forces (on average 10.8% lower than in the linear case) and elongations (on average greater by 19.4%). These changes are smaller than the ones deriving from the variation of other model parameters (e.g. thickness, adhesive energy per unit area). Furthermore, no qualitative changes were observed in the mechanical response of the structures. Therefore, for the sake of simplicity, results are presented here for a linear elastic law. This model correctly reproduces known theoretical results for the delamination of single or multiple tapes (11, 15) or membranes in simple loading cases (16).

For this study, three regions were identified in the elastic membrane: a stiffer fibre (corresponding to the embedded dragline), a circular bridge and a rectangular plaque. The dragline was embedded in the bridge with a variable exit point to perform parametric tests on the centrality parameter $c_d$. This choice was made after analysing the results of parametric tests shown in S1.

These parametric studies show that results could be grouped for small (<45°) and large pulling angles (>120°). In our tests, we thus chose 15°, 90° and 165° as indicative angles to describe small, normal, and large angle loading, respectively (See S1, Figs. S2 and S3). We constructed each a set of models for the anchors of *T. plumipes*, *I. villosa* and *H. troglodytes*, with the specific shape and dragline joint location, using the average geometric variables measured for each species (Tab. 1). Silk membrane stiffness was derived from tensile tests of carefully delaminated silk anchors of *T. plumipes*, *I. villosa* and *H. troglodytes* (9, 12) (further details in S1). Anchors of *K. hibernalis* could not be handled in this setup due to their small size and, accordingly, were omitted. Piriform silk membranes generally had a 10-40 times smaller stiffness than dragline silk of these or related species (17-19) (Tabs. 1, S1). Silk membranes of *T. plumipes* were six times stiffer and stronger than the membranes of *I. villosa* and *H. troglodytes*, on average. This may be due to the grid-like overlay of fibres within the membrane (2) caused by the specific back-and-forth spinning pattern in this spider (10).

A bridge structure was not found in silk anchors of *H. troglodytes*. The effective stiffness of the bridge in *T. plumipes* and *I. villosa* could not be experimentally determined, but based on our structural analysis we presumed a lower stiffness than the plaque, as piriform fibres are barely cross-linked in the bridge, in contrast to the plaque. Furthermore, the bridge contains unstretched, twisted fibres (10) that may contribute to a high extensibility of this structure. We therefore modelled the bridge of *T. plumipes* as 100 times softer than the plaque. The species-specific bridge geometry was modelled based on the microscopy observation of the silk anchors (Tab. 1). In contrast, for *I. villosa* the bridge was modelled as an initially detached circular area around the dragline with the same stiffness as the plaque.

The effect of the positioning of the dragline on the pull-off force is illustrated in detail in the supplementary material, in particular in Figs S1.12.

No model was constructed for *K. hibernalis*, due to the unclear mechanical properties of its silk and the complicated arrangement of dragline fibres in its anchors, which is arduous to implement numerically.

*2.3. Critical role of silk bridge for anchor robustness*

At first, we verified our models by replicating the peeling tests *in silico*. Figures 3g, i and k show the simulated load-displacement curves for different loading angles. For the model of *H. troglodytes*' anchors (Fig. 4a), we observed a decrease in anchor strength for increasing loading angles in agreement with the empirical observation (Fig. 3g). However, we did not observe the decrease in maximal extensibility. This can be attributed to simplifications in the numerical model, which does not account for membrane tearing effects during delamination or a possible anisotropy of the membrane structure.

For the model of *I. villosa*'s anchors, we implemented a detached area centred in the middle of the dragline and with a slightly greater diameter than the dragline length (Fig. 4b). We observed a larger force value and structural stiffness for small pulling angles, whereas similar forces and extensibility were found for loading angles of $\theta = 90°$ and $\theta = 165°$ (Fig. 4i). This improved extensibility for high pulling angles is given by the elongation allowed by the detached area.

For the model of *T. plumipes*' anchors, we implemented a soft bridge centred in a point belonging to the dragline and displaced by 0.08 mm with respect to the centre (Fig. 4c). The diameter of the bridge is 0.5 mm, which is 0.1 mm greater than the dragline length. This causes the rear of the dragline to be connected directly to the plaque only in for a small length, as found in a previous in-depth microscopy study (10). We observed a strong correlation between the structural stiffness and the pulling angle (Fig. 3k). The peak of the force-displacement curves were points of extreme instability of the membrane, corresponding to a snap (i.e. sudden catastrophic failure of the adhesive interface) of the entire structure. Both these phenomena are in agreement with empirical observations.

The analysis of load-displacement curves revealed that, for all species, the force is mostly given by the elastic deformation of the structure. Reaching a border of the membrane causes a drop in the adhesive force, so that the maximal resistance is often found in the early stages of the delamination process. At low pulling angles, (e.g. 15°) the attachments can sustain high maximal loads due to the stiffness of the dragline. At larger pulling angles (90° and 165°), the mechanical response of the structure is ruled by the interaction between the dragline, the bridge and the plaque. For *H. troglodytes* and *I. villosa*, the plaque is pulled along with the dragline, thus lowering the curve slope by the same amount for both 90° and 165°. For 165° the loading force and the dragline are misaligned, lowering the stress concentrations along the dragline and consequently the elastic energy that can be absorbed. This misalignment is reduced if a detached area appears around the dragline, which allows the *I. villosa* anchorage to have similar maximal pull-off forces for 90° and 180°. This changes for the *T. plumipes*, which has a soft bridge surrounding the front of the dragline. For a pull-off angle of 90°, the stresses are distributed from the dragline to the bridge and the plaque, which is directly connected at the rear of the dragline: thus, the structural stiffness is given by the dragline-bridge-plaque system. For a pull-off angle of 165°, the stress is concentrated at the front of the attachment and thus the junction between the dragline and the plaque is not involved in delamination. The stressed part is the soft bridge, thus allowing the dragline to align itself to the loading force, distributing stresses along its length and providing high strength. In conclusion, a soft bridge helps the distribution of stresses along the dragline, thus improving the adhesive performance of the structure at high pulling angles, something which is impossible to achieve by varying other parameters. The variations in effective stiffness for increasing extensions overestimate experimental values. The loading process in the numerical simulation initially involves the soft bridge, with the dragline and the plaque being involved for larger elongation values. In the experiments, a more linear behaviour is observed, with little effective stiffness variation. This would seem to indicate the presence of geometrical features that allow a even more uniform stress redistribution during loading in this specific case, as suggested by our model. To further investigate the dependence of detachment behaviour on geometrical and mechanical characteristics for perpendicular pulling, we varied different parameters in the *T. plumipes* model, like the size and position of the bridge, dragline placement $c_d$ and the ratio between plaque and bridge stiffness $E_b/E_p$ (Tab. 2). We found that for a fixed $E_b/E_p$ the size and position of the bridge had an effect on both the maximal adhesive strength and the slope of the force-extension curve (Fig. 5a). A bridge that was shifted towards the loading point, reflecting the asymmetry of the bridge that is present in *T. plumipes*, could resist the highest force, although the optimal placement depends on many parameters, and is not discussed in detail here. The effect of dragline placement on anchor resistance varied depending on the size and position of the bridge. The empirically observed mechanical behaviour was only obtained in simulations if the bridge surrounded the whole dragline joint and was modelled approximately 100 times softer than the plaque (Fig. 5b). However, this was not the case for more realistic models, where the bridge was positioned asymmetrically with respect to the loading point (Figs. 5c, d). Comparing the latter configurations (*II-IV*) to the one where there is a complete envelopment of the dragline (*I*) we observed that, if a fraction of the dragline is directly connected to the plaque, the stress locations are shifted to the rear plaque (Fig. 5), leading to a better exploitation of the available surface area and higher pull-off forces but a force-displacement slope less dependent from the bridge stiffness. Notice that configuration *III*, where the dragline is also connected to the front plaque, does not lead to a shift in the stress distribution to the rear of the membrane and thus the maximal elongations and pull-off forces are lower than the ones observed in configurations *II* and *IV*.

Focusing now on a single configuration, a softer bridge helps this shift of the stress location to the rear of the membrane, but this does not always lead to a higher force: in configurations *II* (Fig. 5c) and *IV* (Fig. 5d), for $c_d = 0.5$ mm and $E_b/E_p = 100$ lower pull-off force are observed because the rear border is reached sooner that the front one. Independent of the configuration, a softer bridge always increased the maximal elongation of the structure.

## 3. Discussion

We found that the mechanical performance of silk anchors varied profoundly between species, and was highest in the derived orb web spiders. These spiders construct large aerial webs, which requires anchorages that resist delamination independent of the pulling angle (10, 12, 20). It was previously suggested that the structure of silk anchorages was optimized with the evolution of aerial spider webs, which was primarily attributed to the dragline placement parameter $c_d$ (12). The large sheet webs of the ancient *H. troglodytes* were also partly suspended, but were generally anchored with more lines than orb webs and only occurred in sheltered places such as in caves and in hollow logs, and thus the mechanical requirements to their anchors are probably less stringent. Generally, it can be assumed that the requirements to spin robust anchorages are highest in the orb weaver, were webs are exposed to winds, rain and the impact of large prey items. Such impacts must be shared by few anchor lines, each of which are attached with few anchors. Naturally occurring pulling angles may be more steep and more variable in the suspended webs of *T. plumipes* and *H. troglodytes* than in *K. hibernalis* and *I. villosa* that predominantly spin silk lines parallel to the substrate. Both the different selective pressures on anchor robustness and the evolutionary opportunity of anchor enhancement explain the observed inter-species differences in anchor strength and its robustness towards variations in pulling angles.

Our results demonstrated that dragline placement is not the only key parameter governing the strength of silk anchors. In addition, both the ratio of anchor and dragline stiffness as well as the nature of the dragline joint as well as bridge position displayed considerable effects. Most noteworthy, the introduction of a significantly softer silk bridge between the dragline and the plaque heavily enhanced the anchor's capacity to resist detachment, regardless of loading angles. Careful consideration in numerical simulations of numerous other mechanical and geometrical parameters indicates that the presence of such a soft bridge is the most likely mechanism that allows to generate the experimentally observed superior behaviour

(See S1 for a discussion of parametric tests). An indirect attachment through a soft element is thus excellently suited for flexible objects such as threads.

The proposed mechanism of enhanced robustness in anchors with a soft bridge is the differential load sharing between stiff and soft structures, depending on the loading situation. An anchor can sustain higher forces if crack propagation in the adhesive interface is delayed. Once the delamination crack (i.e. the peeling line) reaches one edge of the plaque, the force resistance drops. Thus, not only the size, shape and location of the bridge, but also dragline placement and the ratio between plaque, bridge and dragline stiffness must be well balanced. We assume that of the four species studied here in *T. plumipes* these parameters were better optimized by constant selective pressure on the secure anchorage of large orb webs over the course of hundreds of million years of evolution.

Comparable attachment structures, like the byssus threads of mussels, are more rigid and, accordingly, show a high sensitivity towards the pulling angle (21). Mussels bypass this problem by using not only one byssal attachment but a number of byssus threads that are spaced so that the mean pull-off angle is optimized (22). Notably, the byssus threads exhibit a softer domain in the proximal section, which buffers stress peaks under dynamic loading (22). Material gradients also play a role in the performance of the adhesive structures of beetle feet that are used in highly dynamic situations (23). Silk anchors differ from the previous examples by their rapid production and universal application. Our study demonstrates that the combination of stiff and soft elements significantly enhances the robustness of spider web anchorages.

It is still not fully understood, why the observed relationship between dragline placement $c_d$ and the anchor strength $F_{max}/A$ was expressed more strongly in the two species that build a bridge into their anchors. While it was possible to adjust our numerical models to approximate the empirical force-distance curves, we could not fully replicate a strong effect of $c_d$ on $F_{max}$ in our simulations with bridge models. Such a relationship was only found if the entire dragline joint was surrounded by the bridge, albeit far less expressed than empirically observed. If an asymmetric bridge geometry was used as in the natural anchors, the effect of $c_d$ on $F_{max}$ was negligible (Fig. 5). Also, the effect was expressed more strongly if the difference in stiffness between bridge and plaque was higher. These results indicate that the geometry, size and mechanical properties of the bridge all affect the capacity of dragline placement to enhance anchor strength. Note that a previous set of simulations considering a simpler geometry that neglected the presence of the bridge in the attachment revealed a larger effect of $c_d$ on $F_{max}$

(12). Simulations on this simplified geometry also showed that a central dragline placement had an effect comparable to that of the bridge on the anchor's sensitivity towards pulling angle, but could not explain the differences in the slopes observed in the comparative pulling tests with *T. plumipes*. This shows up the difficulties of adopting an appropriate numerical approach, which must balance between (over-)simplification and (over)-complexity (and thus computation ability). For instance, the natural silk membrane likely has constitutive anisotropic and non-linear properties. This may be particularly relevant in the bridge, where anisotropy has a considerable effect on the stress distribution. Furthermore, fracture was neglected in the models, which may lead to an over-estimation of $F_{max}$ in some instances where anchors would naturally fail by fracture. The balance between adhesion and cohesion failure in thin film adhesive micro-composites such as silk anchors is an important problem that should be clarified in future studies.

## 4. Conclusion

Spiders are masters in the building of web architectures. This can be attributed to the combination of different materials and structures in web anchorage. We have demonstrated that robust attachments are achieved by using a material that is significantly softer than the structural line. The introduction of an additional softer element between the substrate attachment and the line prevents an early failure at steep loading angles, and enhances the versatility of the anchor. This is most pronounced in orb weavers, where the structure and the combination of extensible and stiffer domains are optimized to create a robust system.

From the engineering perspective, spider web anchorages could be a truly inspirational model to design nearly instant, durable and traceless attachment systems that do not require an alteration of the substrate, with a multitude of applications, such as adhesive joints of components in vehicle construction, the flexible fastening of textile roof structures in architecture, safety line anchorages in industrial and recreational climbing, and attaching items to walls in households. Here we have demonstrated the benefit of a two-domain or graded adhesive joint, a principle that it is worth exploring in future applied research.

## 5. Material and Methods

*Material sourcing and fieldwork*

Spiders were collected in and around Sydney, New South Wales (*T. plumipes*, *I. villosa*) and Dover, Tasmania (*H. troglodytes*) on public parks or private properties with landlord permission. Samples from *K. hibernalis* were collected from a lab stock at the Technical University of Aachen. Silk anchors were collected on poly-propylene sheets (cut pieces of document protector sheets) that were deposited in the spider's containers and removed after 2-5 days. Single anchors were isolated by cutting the polymer sheet ~1 cm around the sample and carefully cutting or burning (with hot needles) the attached dragline ~2 cm above the membrane. For mounting into the test setup, the polymer sheet with the attached silk anchor was attached to a microscopy glass slide using double-sided tape.

*Structure and morphometrics of silk anchors*

Nine to twenty silk anchors per individual spider were imaged with Leica M205A (Leica Microsystems GmbH, Wetzlar, Germany) and Motic (Motic Inc. Ltd., Hong Kong) stereo microscopes with mounted cameras.

Morphometrics of silk anchors was performed on micrographs in *ImageJ* (24). We calculated the dragline placement variable $c_d$ as follows: distance $d$ between the dragline joint (point were the dragline leaves the anchor) and the anterior border of the anchor divided by the longitudinal dimension of the anchor. Details on the morphometric characterization of silk anchors are described in ref. (10).

*Silk biomechanics*

We measured the maximal force ($F_{max}$) required to pull off a silk anchor from a polypropylene surface as described in detail in ref. (10). Tests were performed under dry lab conditions (~50% relative humidity, 22-26°C). Each silk sample was photographed prior to measurement to determine $c_d$ and $A$ (see above). Primary data was obtained for wild-caught *Hickmania troglodytes* and lab-reared *Kukulcania hibernalis*. Primary data for *Trichonephila plumipes* and *Isopeda villosa* was taken from refs. (10) and (20) and re-analysed. Data of different individuals was pooled for statistical analysis. $F_{max}/A$ was statistically compared between species using pairwise Wilcoxon rank sum tests with FDR alpha correction. To explore the correlation of $c_d$ and $F_{max}/A$, we fitted sine and linear functions. While the relationship will likely follow a sine function, linear functions provided a similarly good fit (S2), because data for very low and high $c_d$ was sparse or lacking, since they are rarely produced by spiders. We therefore proceeded with linear regression using R 3.5.0. (25).

*Numerical model*

The elastic membrane was modelled by discretising it in a network of elastic bonds (i.e. springs) in a square-diagonal lattice, using a generalized non-linear 3D co-rotational truss formulation(26), as done in Liprandi et al. (14). A homogenization procedure was adopted, imposing the equivalence of the strain energy density of the lattice with that of a corresponding homogeneous membrane (27, 28). The interface was modelled using a Cohesive Zone Model characterized by a 3D traction-separation law $T_i = \Delta_i \frac{\phi}{\delta^2} \cdot \exp\left(-\frac{\Delta_{eff}^2}{\delta^2}\right)$ where $\phi$, $\Delta_{eff}$ and $\delta$ are the work of separation, the crack gap value and the characteristic length (i.e. the gap value corresponding to the maximum traction) (29, 30).

A displacement-control procedure was used to simulate the external load. The resulting system of coupled non-linear equations in matrix form was solved using an algorithm based on the Newton-Raphson method (31) implemented in C++. The choice of the crack gap value was made, as already done in literature, to be of the order of the discretization length and consistent with analytical results for axisymmetrical peeling.

Two model parameters were calculated using the available empirical data (Tabs. 1, S1). Since the model represents a continuous membrane, we took into account the sparseness of the fibres by recalculating the elastic coefficient $E \cdot t$ by comparing the force-displacement relationship of the numerical model and the empirical results during the purely elastic phase (i.e. before any delamination occurs, and thus without the involvement of adhesive energy). We supposed that during the elastic phase the anchorages would behave as an elastic circular membrane (16) with a Young's modulus between those of the bridge and dragline. Then, we calculated the thickness needed to obtain a force-elongation ratio equal to the empirical one, obtaining values 10 to 100 times lower than the initially estimated ones for all the species. We thus chose to use a thickness equal to 0.1 times the empirical value (see Results section). The work of separation $\phi$ was then chosen for each silk anchor by comparing the maximal forces and displacements obtained by the numerical simulations with the empirical ones using a trial-and-error scheme for a pull-off angle of $\theta = 90°$ (where the highest proportion of samples failed by total delamination in this loading situation in all three species, allowing us to neglect fracture). We obtained $\phi = 4$ kPa/mm for the *H. troglodytes* and the *I. villosa*, and $\phi = 2$ kPa/mm for the *T. plumipes*. The morphology of the anchors was simplified in the models. For the plaque, a rectangular shape was chosen with the dimensions taken as the length and width measured in the samples (see above). In *H. troglodytes* the dragline was model as directly fused with the plaque. In the *I. villosa* model a circular bridge was introduced around the fused dragline, using the same Young's modulus as the plaque (0.22 GPa). In the *T. plumipes* model the bridge centre was positioned closer to the loading point and assigned a Young's modulus of 0.016 GPa, which was 100 times lower than the plaque to simulate the higher extensibility provided by the unstretched bridge fibres (see main text for justification). The exact parameters used are given in Tab. 1 and schematically illustrated in Fig. 6.

The pulling procedure is performed as follows: a fixed small displacement is applied at the insertion point at an angle $\theta$. The equilibrium configuration is obtained in quasi-static conditions using an iterative procedure. Another fixed small displacement is then applied in the same point at the same angle, and so forth. The simulation stops when one of the following becomes true: (a) total delamination is reached; (b) a mechanical instability resulting in a snapping of a part of the membrane is reached.

To observe how the geometrical and mechanical properties change the mechanical response of the elastic membrane, two explorative parametrical study were conducted. We studied the effect of the dragline width, the dragline length, the adhesive energy per area, the anchorage thickness and the stress-strain relationship of the silk for a disc with of an area of 1.4 mm x 1.6 mm plaque, with a dragline of length equal to 0.4 mm inserted at 0.5 mm from the front border. The results are shown in S1. For the *T. plumipes* model, we studied the geometrical and mechanical parameters of the bridge by varying the dragline attachment and bridge shapes as follows: three different centres of the bridge were used: at the insertion point $c_d$, at half the dragline length $(c_d + \frac{l_d}{2})$, and close to the half of the dragline length $(c_d + \frac{l_d}{2} - \Delta b)$; and four different bridge radii (Tab. 2). The bridge Young's modulus $E_b$ was set to be $E_b = 0.1\, E_p$ or $E_b = 0.01\, E_p$, where $E_p$ is the Young's modulus of the plaque. Three different values of $c_d$ ($[0, 0.25, 0.5]$ mm) were used for each bridge configuration.

**Acknowledgements**

We are grateful to Arthur Clarke and Niall Doran, who helped with the collection of Tasmanian cave spiders, and to Gabriele Greco for useful discussions.

**Funding**

This study was supported by a Macquarie University Research Fellowship of Macquarie University and a Discovery Early Career Researcher Award of the Australian Research Council (DE190101338) to JOW. FB and NMP are supported by the European Commission under the H2020 FET Open ("Boheme") grant No. 863179. DL, FB and NMP were supported by COST Action CA15216 "European Network of Bioadhesion Expertise". ACJ was funded by the Deutsche Forschungsgemeinschaft (JO1464/2-1). NMP was supported by the Italian Ministry of Education, University and Research (MIUR) under the "Departments of Excellence" grant L.232/2016, the ARS01-01384-PROSCAN Grant and the PRIN-20177TTP3S Grant.

**Author contributions**

JOW, FB and NMP devised, led and administered the project. JOW and ACJ performed the empirical study. DL developed the numerical model and carried out the numerical simulations. JOW, DL and FB wrote the manuscript and NMP reviewed it. All authors critically revised the manuscript and approved its final version.

**Competing Interests**

The authors declare that they have no conflict of interests.

**Data accessibility**

Data are reported in the supplemental information (S1, S2) and the main text of this paper.

# Tables

**Tab. 1. Geometrical properties of models used in numerical simulations.** For a schematic illustration of the geometrical parameters see Fig. 6a. Abbreviations: $b$, centre of circular bridge $(x, y)$; $d$, location of dragline insertion point (distance from front edge); $E_b$, Young's modulus of the plaque; $E_p$, Young's modulus of the plaque; $E_d$, Young's modulus of the dragline; $l_p$, plaque (membrane) length; $l_{dl}$, length of attached dragline; $r_b$, radius of the circular bridge; $t_{num}$, membrane thickness; $t_{num}/t_{emp}$, relative density (fibre density); $w_p$, maximal width of plaque (membrane).

| | $l_p$ (mm) | $w_p$ (mm) | $l_{dl}$ (mm) | $d$ (mm) | $E_p$ (GPa) | $E_d$ (GPa) | $E_b$ (GPa) | $b$ (mm) | $r_b$ (mm) | $t_{num}$ (µm) | $t_{num}/t_{emp}$ |
|---|---|---|---|---|---|---|---|---|---|---|---|
| *H. troglodytes* | 2.4 | 2.6 | 1.2 | 0.5 | 0.25 | 10 | - | - | - | 1 | 0.1 |
| *I. villosa* | 2.1 | 1.6 | 1.3 | 0.4 | 0.22 | 10 | 0.22 | (1.0, 0.8) | 0.7 | 1 | 0.1 |
| *T. plumipes* | 1.4 | 1.5 | 0.4 | 0.5 | 1.68 | 15 | 0.0168 | (0.62, 0.75) | 0.25 | 1.5 | 0.1 |

**Tab. 2. Properties of models used in the parametrical study on the effect of bridge geometry and softness on pull-off forces.** For a schematic illustration of the geometrical parameters see Fig. 6a. Abbreviations: $b$, centre of circular bridge; $E_p$, Young's modulus of the plaque; $E_d$, Young's modulus of the dragline; $l_p$, plaque (membrane) length; $l_{dl}$, length of attached dragline; $r_b$, radius of the circular bridge; $t_{num}$, membrane thickness; $w_p$, maximal width of plaque (membrane). $d$ varied for every set of simulations (I-IV) between the following values: 0 mm, 0.25 mm, 0.5 mm. $E_b$ varied for every set of simulations (I-IV) between the following values: 0.168 GPa, 0.0168 GPa.

| | $l_p$ (mm) | $w_p$ (mm) | $l_{dl}$ (mm) | $E_p$ (GPa) | $E_d$ (GPa) | $t_{num}$ (µm) | $b$ (mm) | $r_b$ (mm) |
|---|---|---|---|---|---|---|---|---|
| I | 1.4 | 1.5 | 0.4 | 1.68 | 15 | 1.5 | $d + 0.2$ | 0.25 |
| II | 1.4 | 1.5 | 0.4 | 1.68 | 15 | 1.5 | $d$ | 0.2 |
| III | 1.4 | 1.5 | 0.4 | 1.68 | 15 | 1.5 | $d + 0.12$ | 0.125 |
| IV | 1.4 | 1.5 | 0.4 | 1.68 | 15 | 1.5 | $d + 0.12$ | 0.25 |

**Figure legends**

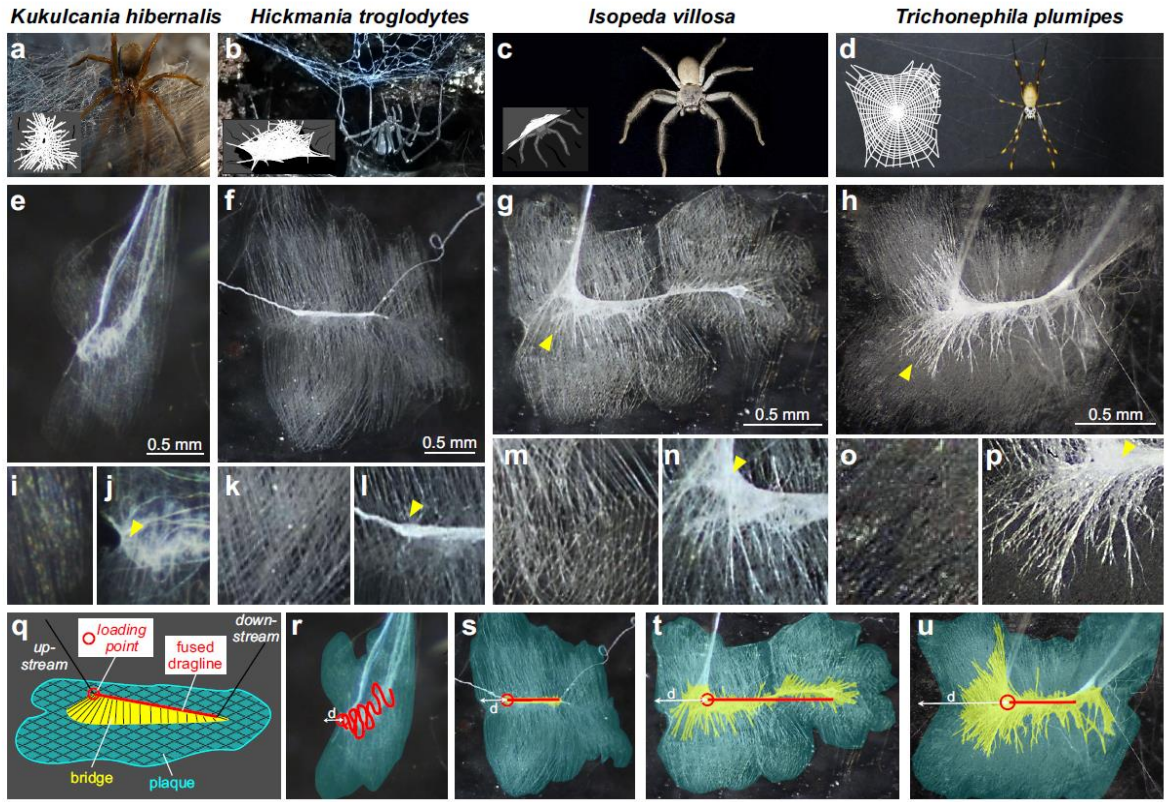

**Fig. 1. Morphology of spider silk line anchors**. (a, e, i, j, r) Southern house spider (*Kukulkania hibernalis*, Filistatidae). (b, f, k, l, s) Tasmanian cave spider (*Hickmania troglodytes*, Austrochilidae). (c, g, m, n, t) Huntsman (*Isopeda villosa,* Sparassidae). (d, h, o, p, u) Golden orb weaver (*Trichonephila plumipes*, Araneidae). (a-d) Photos of spider species used in the tests, and for which anchors are displayed in columns below. Schematic insets display web or shelter architecture. Note that a-c are substrate bound silk structures with anchor lines meeting the substrate at low angles, whereas d is an aerial (suspended) web with anchor lines meeting the substrate at a steep angle. (e-h) Total view of silk anchor on a transparent substrate, viewed from above in a stereo microscope with reflective light. Yellow arrowhead points to the 'bridge' element. (i, k, m, o) Detail of plaque. Note the lattice-like arrangement of piriform silk fibres and the higher fibre density in *T. plumipes*. (j, l, n, p) Detail of dragline joint. Yellow arrowhead points to the loading point. Note the piriform fibre network (bridge) in n and p. (q) Schematic illustration of silk anchor as viewed at an angle. (r-u) same as in e-h but with the different elements marked (same colour code as in q): red, fused dragline (schematically); blue, plaque; yellow, bridge; red circle, loading point (point where the dragline leaves the anchor); $d$, distance between the loading point and the plaque edge (dragline placement $c_d = d/l_p$, where $l_p$ is the plaque length).

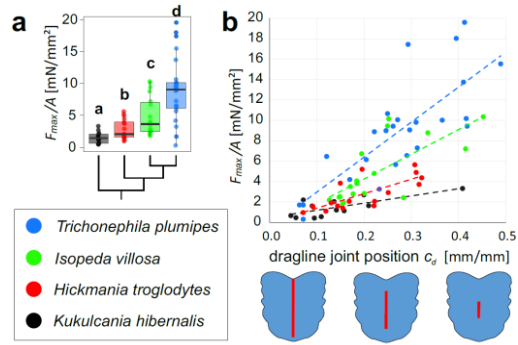

**Fig. 2. Comparison of web anchor performance.** (a) Measured anchor strength $F_{max}/A$ for perpendicular pulling ($\theta = 90°$) in four species of spiders. Below, a simplified cladogram of the species relationships. Different letters above the plots indicate different groups with a significance level of 0.05. (b) $F_{max}/A$ against dragline placement $c_d$ for each of the species (same colour code as in a). Schematic drawings below the diagram illustrate silk anchors with different $c_d$ values (as viewed from above) according to the diagram's x-axis (fused dragline in red, piriform silk in blue).

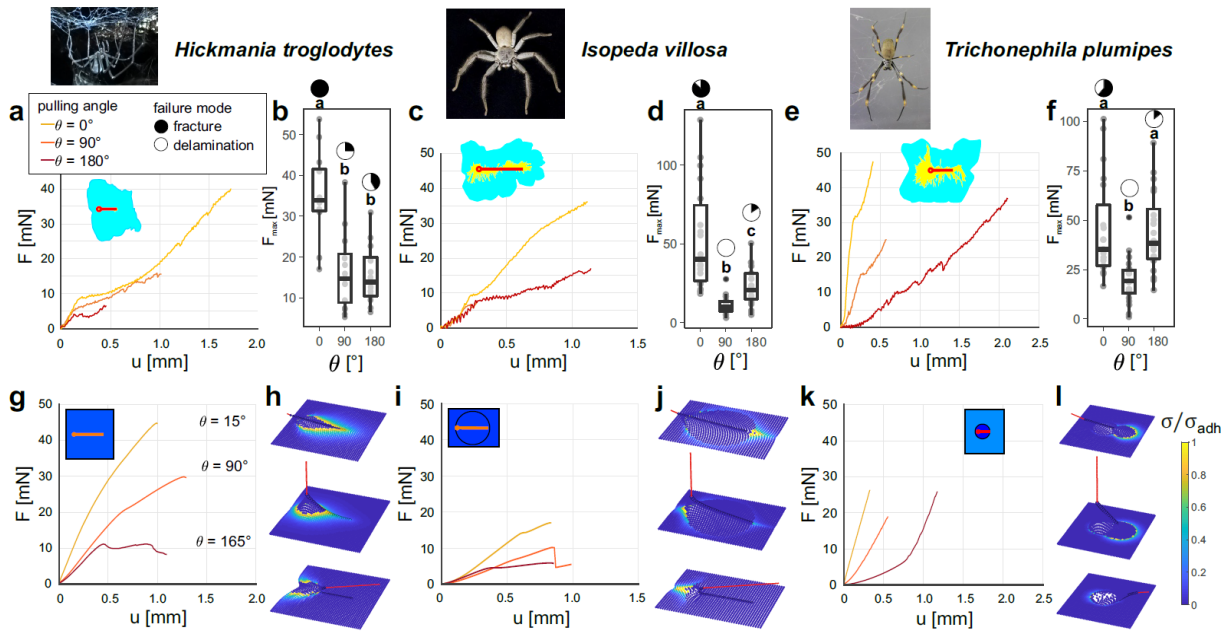

**Fig. 3. Pull-off dynamics of silk anchors in three spider species.** (a-f) Results of empirical pull-off tests. (g-l) Results of numerical simulations. (a-b, g-h) Tasmanian cave spider (*H. troglodytes*). (c-d, i-j) Huntsman (*I. villosa*). (e-f, k-l) Golden orb weaver (*T. plumipes*). (a, c, e) Exemplary force-extension curves for three different pulling angles ($\theta$). Note that the initial slope that indicates the stiffness of the structure, is similar between loading situations in *H. troglodytes* and *I. villosa*, but decreases with increasing pulling angles in *T. plumipes*. Inset schematics illustrates anchor morphology as viewed from above (see Fig. 1 for details), not to scale. Different letters above the plots indicate different groups with a significance level of 0.05. (b, e, h) Maximal pull-off forces ($F_{max}$) depending on dragline pulling angle. Boxplots indicate the median (thick middle line), the interquartile range (the boxes), extreme values (whiskers) and out-of-range outliers (dark dots), with all original data points in dark grey (jitter plots). Pie charts above boxes illustrate the proportion of samples that failed by rupture (black) versus

complete surface detachment (white). (g, i, k) Simulated force-extension curves with numerical models using the parameters measured in the natural anchors. Colour code of the curves for different pulling angles as in a, c, e. Inset schematics illustrates view of model from above (see Fig. 4 for details), not to scale. (h, j, l) Three dimensional rendering of models at an intermediate displacement ($u/u_{max} \cong 0.8$) at $\theta = 15°$, $\theta = 90°$ and $\theta = 165°$ (from top to bottom). The colours indicate the relative adhesive stress field ($\sigma/\sigma_{adh}$), with blue indicating low stress and yellow maximal stress ($\sigma_{adh}$) causing delamination (see legend right of l). The red line illustrates the pulled upstream dragline.

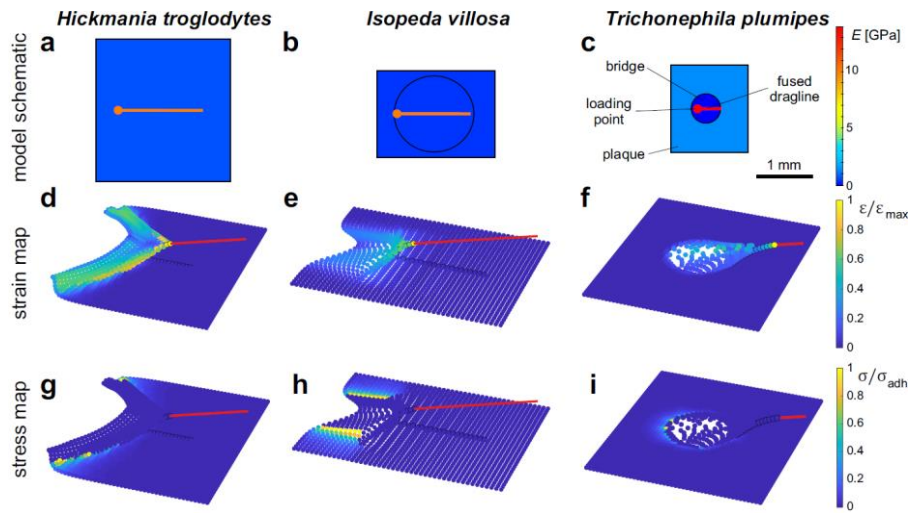

**Fig. 4. Simulated behaviour of silk anchor under a steep loading angle.** Species models grouped per column. (a-c) Schematic illustration of model, as viewed from above, all same scale. Colours indicate the stiffness (E) of each element (see legend to the right). (d-i) Three-dimensional renderings of models at $F_{max}$ loaded at $\theta = 165°$. (d-f) Colours indicate the relative strain field ($\varepsilon/\varepsilon_{max}$), with $\varepsilon$ indicating the mean strain of the springs representing a section of the membrane, $\varepsilon_{max}$ indicating the maximal strain at the correspondent timestep, blue colours indicating low strain and yellow indicating high strain (see legend to the right). (g-i) Colours indicate the adhesive stress field ($\sigma/\sigma_{adh}$), with $\sigma$ indicating the stress in the springs at the adhesive interface, $\sigma_{adh}$ indicating the stress at which delamination occurs, blue colours indicating low stress and yellow indicating high stress (see legend to the right). The anchors of *H. troglodytes* and *I. villosa* fail by delamination (peeling), whereas the anchor of *N. plumipes* detaches by a sudden catastrophic failure ('snap') of the adhesive interface.

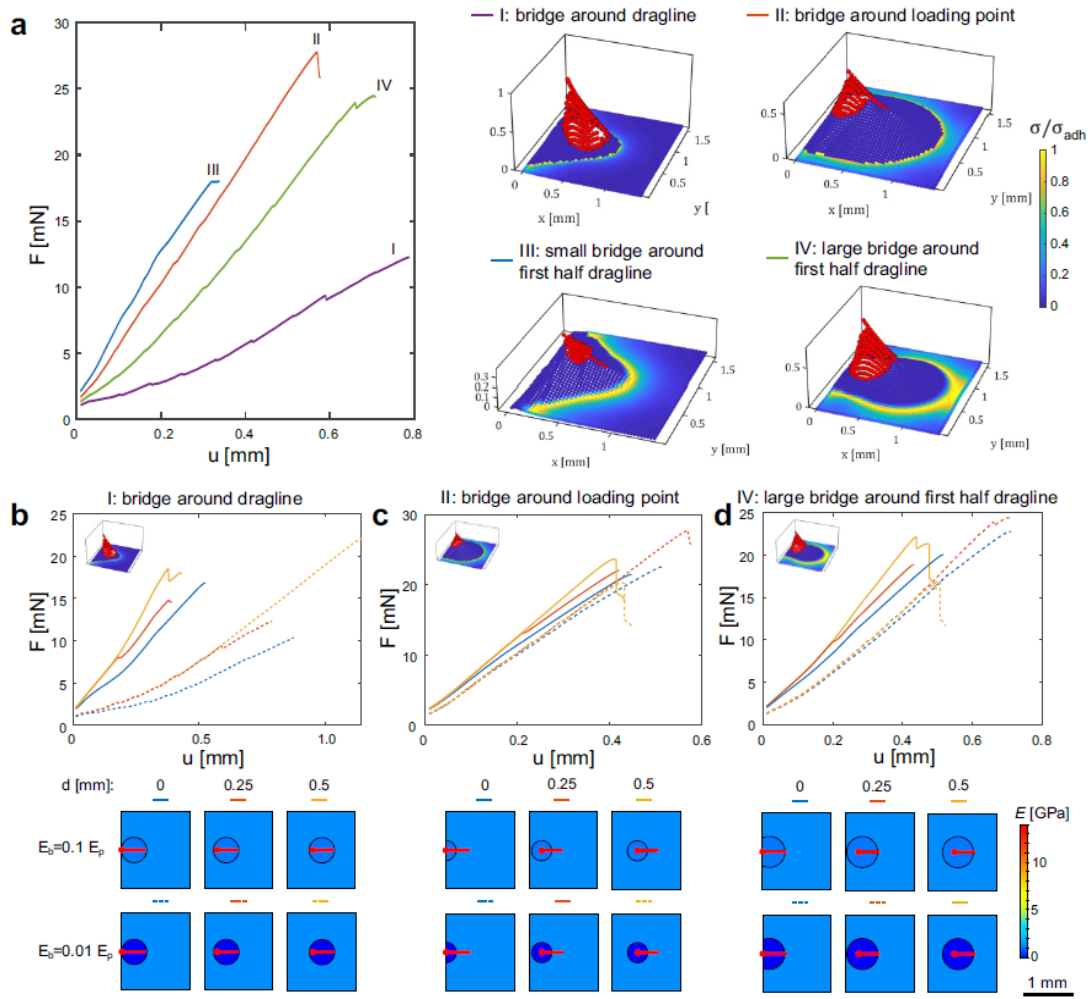

**Fig. 5. Effect of bridge structure and stiffness, and dragline placement on anchor resistance.** (a) Simulated force-extension for a $\theta = 90°$ using different bridge geometries. Three-dimensional renderings on the right illustrate the different models under load, with red illustrating the bridge and dragline joint, and blue and yellow shading indicating interfacial stress in the plaque (peeling line). (b-d) Simulated force extension curves for varying dragline placement $c_d$ and bridge softness. Solid lines represent the force-extension curves of models with $E_b = 0.1\ E_p$ and dashed lines of models with $E_b = 0.01\ E_p$ (i.e. softer bridge). The morphology and mechanical properties of the corresponding models are schematically illustrated below, as viewed from above (same conventions as in Fig. 4).

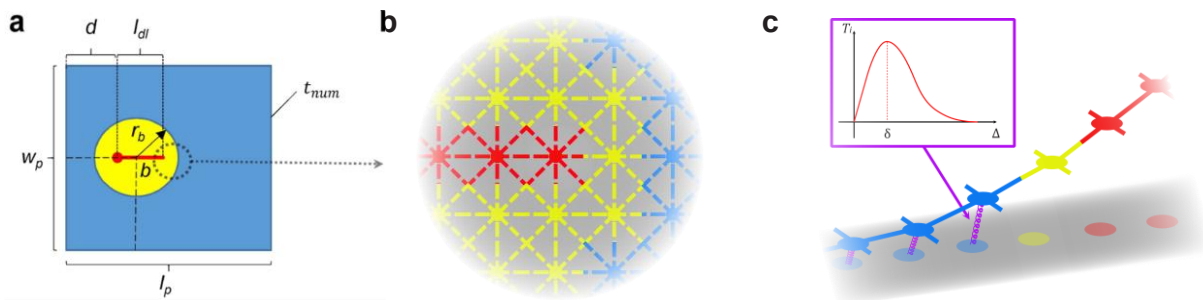

**Fig. 6. Schematic illustration of properties simulated in the numerical models.** The attached plaque is coloured blue, the bridge (initially detached, softer part of plaque) in yellow, and the fused part of the dragline in red. Different elastic and adhesive properties were assigned to each of these three parts. (a) Geometrical parameters of anchor models. (b) Schematic representation of the Lattice Spring Model (LSM). Elastic springs connect the nodes. Springs connecting two different areas have a stiffness equal to the mean of the two assigned values. (c) Schematic representation of the cohesive model. Each plaque node is connected to its initial position on the substrate by a traction-separation law.

**Supplemental information**

S1. Results of parametric studies and information on the inference of mechanical properties of silk anchor parts for in silico experiments (pdf).

S2. $F_{max}$ and $F_a$ data for four species of spiders (csv).